\documentclass[prl,twocolumn,nopacs,preprintnumbers,notitlepage,amsmath,amssymb,superscriptaddress]{revtex4-2}

\usepackage{graphicx}
\usepackage[utf8]{inputenc}

\begin{document}

\def\be{\begin{equation}}
\def\ee{\end{equation}}
\def\bea{\begin{eqnarray}}
\def\eea{\end{eqnarray}}

\title{A positive answer on the existence of correlations between positive earthquake magnitude differences}

\author{Eugenio Lippiello}
\affiliation{Department of Mathematics and Physics, University of Campania “Luigi Vanvitelli”, Caserta, Italy}
\author{Lucilla de Arcangelis}
\affiliation{Department of Mathematics and Physics, University of Campania “Luigi Vanvitelli”, Caserta, Italy}
\author{Cataldo Godano}
\affiliation{Department of Mathematics and Physics, University of Campania “Luigi Vanvitelli”, Caserta, Italy}

\date{\today}
\begin{abstract}
  The identification of patterns in space, time, and magnitude, which could potentially encode the subsequent earthquake magnitude, represents a significant challenge in earthquake forecasting. A pivotal aspect of this endeavor involves the search for correlations between earthquake magnitudes, a task greatly hindered by the incompleteness of instrumental catalogs.
A novel strategy to address this challenge is provided by the groundbreaking observation by Van der Elst (Journal of Geophysical Research:
Solid Earth, 2021), that positive magnitude differences, under certain conditions, remain unaffected by incompleteness. In this letter we adopt this strategy which provides a clear and unambiguous proof regarding the existence of correlations between subsequent positive magnitude differences. Our results are consistent with a time-dependent $b$-value in the Gutenberg-Richter law, significantly enhancing existing models for seismic forecasting.

\end{abstract}
\maketitle

Seismic catalogs suffer from significant incompleteness. Small earthquakes often go unreported due to their occurrence at distances beyond the detection range of monitoring stations \cite{SW08,MWWCW11,MW12}, or they may be overshadowed by the coda-wave of preceding larger earthquakes \cite{Kag04,HKJ06,PVIH07,LCGPK16,Hai16,Hai16a,dAGL18,PLLR20,Hai21}. This latter mechanism, in particular, renders the catalog inherently incomplete \cite{LCGPK16,LPGTPK19} and has left fundamental questions about seismic activity unanswered.
One such crucial inquiry involves seismic forecasting and the potential dependence of earthquake magnitudes on the organization of seismic events in time, space, and magnitude preceding its occurrence. Studies have convincingly demonstrated correlations between earthquake magnitudes, with these correlations depending on both temporal and spatial distances \cite{LGdA07,LBGD07,LdAG08,SYTR05,VSTS05, SSV09, SSV10, Sar11,LGdA12}. However, the challenge lies in discerning whether these observed correlations are genuine features or spurious artifacts of catalog incompleteness \cite{DG11,LGdA12,NS14,dAGGL16,SS16,GRTVEW18,OKTH18,NOS19,NOS22,PZ22,PZ23,XBG23,Tar24}. Consequently, the most commonly adopted statistical forecasting models, such as the ETAS model \cite{Oga85,Oga88,Oga88b,Oga89}, assume earthquake magnitudes to be independent and identically distributed (i.i.d.) variables. Within this framework, forecasting is severely limited, relying solely on the information that large earthquakes have a higher probability of occurring when and where the seismic rate is higher. Modifications of the ETAS model, incorporating magnitude correlations have been proposed in the literature \cite{LGdA07,SS16b,NOS19,NOS22} 

Recently, Van der Elst \cite{VdE21} introduced a novel and remarkable tool designed to address issues associated with catalog incompleteness in a simple yet highly effective manner. This tool is grounded in the observation that, under specific conditions, positive magnitude differences remain largely unaffected by incompleteness \cite{VdE21,LP24}. Leveraging this approach, researchers have been able to derive the true magnitude distribution, the true occurrence rate \cite{VdEP23}, and the correct parameters of the ETAS model \cite{VdE23}, even when working with incomplete catalogs. In this study, we employ this tool to unequivocally demonstrate the presence of genuine correlations between earthquake magnitudes.

A seismic catalog is a multi-variate dataset ${\cal C}$ consisting of $N$ elements, where the $i$-th element is the vector ${m_i, t_i, \vec{r}_i}$ representing the earthquake magnitude, occurrence time, and epicentral coordinates, respectively, of the $i$-th earthquake. Specifically, we examine the relocated catalog for Southern California (SC) \cite{HYS12}, spanning from January 1981 to March 2022, encompassing $N=800499$ earthquakes with $m \ge -1$. Additionally, we consider the relocated catalog for Northern California (NC) \cite{WS08}, covering the period from January 1984 to December 2021, including $N=879547$ earthquakes with $m \ge -0.8$.
From the original catalog, we also construct a catalog of magnitude differences, denoted as ${\cal D}$, where each element $i$ is represented by the vector ${\delta m_i, t_i, \vec{r}_i}$, with $\delta m_i = m_{i+1} - m_i$ representing the magnitude difference between two successive earthquakes in the original catalog.

To investigate correlations between earthquake magnitudes, we extend the method developed in \cite{LdAG08}. We define the conditional probability as:

\be
P(\delta m_i > M \vert \delta y_i < Y) \equiv \frac{N(M,Y)}{N(Y)}.
\label{pcond}
\ee

Here, $\delta y_i$ denotes either the epicentral distance $\delta r_i = \vert \vec{r}_{i+1} - \vec{r}_i \vert$ or the temporal distance $\delta t_i = t_{i+1} - t_i$, depending on the context. In Eq.(\ref{pcond}), $N(M,Y)$ represents the number of pairs of subsequent events where both $\delta m_i > M$ and $\delta y_i < Y$, while $N(Y)$ is the total number of pairs with $\delta y_i < Y$.
Starting from the initial catalog $\mathcal{C}$, we generate a catalog with reshuffled magnitude differences $\mathcal{D}_{\text{ran}}$, where the $i$-th element is the vector ${\delta m^*_i, t_i, \vec r_i}$, with $\delta m^*_i = m_k - m_i$, and $k$ represents the index of an earthquake randomly chosen within the catalog. Specifically, the catalog $\mathcal{D}_{\text{ran}}$ retains the same occurrence time and epicentral coordinates as catalog $\mathcal{D}$, but the magnitude differences are randomized. From a single instrumental catalog $\mathcal{C}$, we can derive multiple reshuffled catalogs $\mathcal{D}_{\text{ran}}$. In each reshuffled catalog, we evaluate the quantity $P(\delta m^*_i >  M \vert \Delta y_i < Y)$, which exhibits a different value in each reshuffled catalog $\mathcal{D}_{\text{ran}}$. This quantity follows a Gaussian distribution with a mean value $\overline{P(M,Y)}$ and a standard deviation $\Sigma(M,Y)$.

\begin{figure}[htp]
    \centering
    \includegraphics[width=9cm]{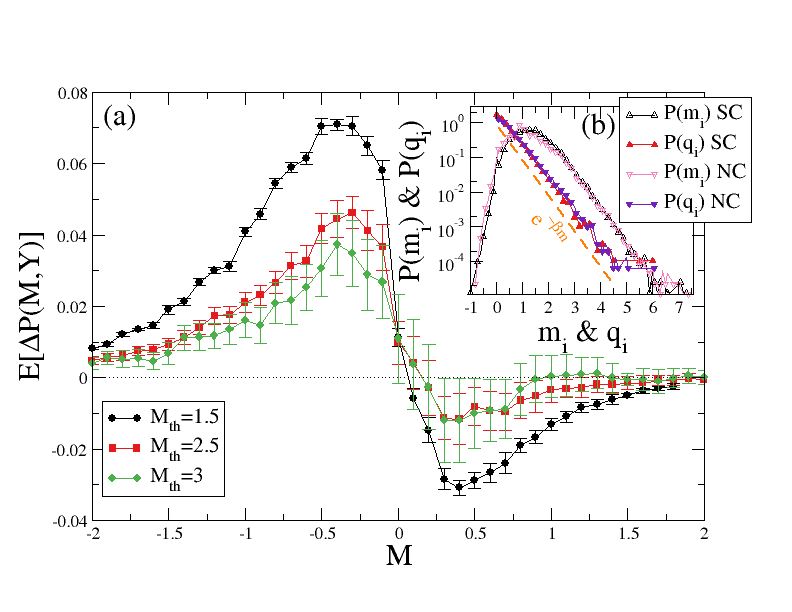}
    \caption{(a) The quantity $E\left[\Delta P(M,Y)\right]$ for SC, where $\Delta P(M,Y)$ is defined in Eq.(\ref{DeltaP}),  with $\delta y_i=\delta r_i$ and $Y=10$ km. Different curves correspond to different thresholds $M_{th}$. Error bars represent $2\Sigma(M,Y)$. (b) The magnitude distribution $P(m_i)$ and the positive magnitude difference distribution $P(q_i)$ for SC and NC. The dashed orange line represents the GR law for SC with $\beta=2.23$.}
    \label{fig1}
\end{figure}

The key quantity of the study is defined as:
\be
\Delta P(M,Y) = P(\delta m_i> M \vert \delta y_i<Y)-P(\delta m^*_i> M \vert \Delta y_i<Y).
\label{DeltaP}
\ee
We then calculate the mean value $E\left[\Delta P(M,Y)\right]$, obtained by averaging over 1000 realizations of the reshuffled catalog $\mathcal{D}_{\text{ran}}$. $E\left[\Delta P(M,Y)\right]$ for SC is plotted in Fig.\ref{fig1}a for the case where $\delta y_i=\delta r_i$ and $Y=10$ km. The error bars represent twice the standard deviation $\Sigma(\delta M,Y)$. Values such as $\vert E\left[ \Delta P(M,Y)\right]\vert>2\Sigma(M,Y)$ indicate that it is highly improbable for a catalog with uncorrelated magnitudes, such as $\mathcal{D}_{\text{ran}}$, to produce the same probability of observing a given magnitude difference $P(\delta m_i> M \vert \Delta y_i<Y)$ of the real catalog. 
Therefore, Fig.\ref{fig1}a demonstrates that magnitudes in the instrumental catalog $\mathcal{C}$ are correlated.
In Fig.\ref{fig1}a, we also examine the influence of a magnitude threshold $M_{\text{th}}$ on $E\left[\Delta P(M,Y)\right]$. By considering only earthquakes with magnitudes larger than $M_{\text{th}}$ and increasing $M_{\text{th}}$, we observe a decrease in $E\left[\Delta P(M,Y)\right]$. This decrease can be interpreted as a signature that magnitude correlations are primarily a spurious effect of incompleteness, as increasing $M_{\text{th}}$ focuses on a catalog less affected by incompleteness. Nevertheless, the observation that even for $M_{\text{th}} = 3$, which exceeds the completeness magnitude $M_c$ estimated from the magnitude distribution (Fig.\ref{fig1}b), $\vert E\left[\Delta P(M,Y)\right]\vert>2\Sigma(M,Y)$ suggests the persistence of genuine magnitude correlations beyond incompleteness. This scenario finds further support in the resilience of $E\left[\Delta P(M,Y)\right]$ to variations in the quality of the seismic network \cite{LGdA12} and is reinforced by laboratory experiments on rock fracture \cite{XBG23}. These experimental findings demonstrate the existence of magnitude clustering irrespective of loading protocols, rock types, and observable magnitude ranges.

Next, we will review these results in light of recent findings concerning positive magnitude difference statistics.
We assume as null hypothesis that magnitudes are i.i.d variables which obey the Gutenberg-Richter (GR) law, $p(m) \propto \exp(-\beta (m-M_L))$  for magnitudes larger than a lower limit $M_L$. Although a precise estimate of $M_L$ is unavailable, it is reasonable to assume that $M_L$ takes on a very negative value \cite{dAGGL16}. Additionally, we  assume that not all magnitudes are reported in the catalog due to detection issues. Therefore, the observed magnitude distribution is given by \cite{LP24}
 \be
 p\left (m_i\right) =\beta e^{-\beta (m_i-M_L)}
 \Phi\left(m_i-M_c\right),
 \label{punfac}
 \ee
 where $M_c>M_L$ is the completeness magnitude and $\Phi(x)$ is a detection function which is a monotonic increasing function of $x$ ranging from $\Phi(x)=0$ for
 $x \lesssim -2\sigma$ to $\Phi(x)=1$ for $x \gtrsim 2\sigma$. Essentially, earthquakes with magnitudes $m_i<M_c-2\sigma$ are undetected, while those with $m_i>M_c+2\sigma$ are always detected. A realistic functional form for $\Phi(x)$ is $\Phi(x)=\left(1+\text{Erf}(x/\sigma)\right)/2$, with typical estimates placing $\sigma$ in the range $\sigma \in (0.2,0.3)$ \cite{OK93,VdE21,LP24}.
Eq.(\ref{punfac}) is consistent with the magnitude distribution of instrumental catalogs (Fig.\ref{fig1}b), which conforms to the GR law only for magnitudes $m_i>M_c$, determined using the maximum curvature method  \cite{WW05}, with $M_c=2.6$ for SC and $M_c=2.7$ for NC (Fig.\ref{fig1}b).

Using Eq.(\ref{punfac}) in Eq.(\ref{pcond}), we obtain
\be
\begin{aligned}
& P(\delta m_i>M \vert \Delta y_i<Y)= \beta^2 \int_{m_L}^{\infty}dm_i \int_{m_i+M }^{\infty}dm_{i+1} \\
& e^{-\beta (m_{i+1}+m_i-2 m_L)} \Phi\left(m_i-M_c\right) \Phi\left(m_{i+1}-M_c\vert m_i\right). 
\end{aligned}
\label{pcond2}
\ee
In the above equation, we explicitly use the notation $\Phi\left( m_{i+1}-M_c\vert m_i\right)$ to specify that the detection function must be evaluated under conditions where the previous earthquake $m_i$ has been identified and reported in the catalog.
It is worth noticing that it is exactly this conditioned detection function that introduces correlations between the magnitudes $m_i$ and $m_{i+1}$. Indeed, the same expression Eq.(\ref{pcond2}) holds for the probability $P(\delta m^*_i> M \vert \Delta y_i<Y)$, with $m_k$ instead of $m_{i+1}$.
This makes a fundamental difference since, by construction, $k$ is a random index, and therefore $\Phi\left(m_{k}-M_c\vert m_i\right)=\Phi\left(m_{k}-M_c\right)$, independently of $m_i$.

Accordingly, $\Delta P(M,Y)$ can be rewritten as
\be
\begin{aligned}
& \Delta P(M \vert Y)= \beta^2 \int_{m_L}^{\infty}dm_i \int_{m_i+M }^{\infty}dm_{i+1} 
 e^{-\beta (m_{i+1}+m_i-2 m_L)} \\ & \Phi\left(m_i-M_c\right)
\left [\Phi\left(m_j-M_c\vert m_i\right) -\Phi\left(m_j-M_c\right)\right].
\end{aligned}
\label{dp}
\ee
The origin of magnitude correlations in the null hypothesis scenario originates from the difference of the two detection functions in the term between square brackets in Eq.(\ref{dp}). It is evident that, if $M >0$, $m_j>m_i$ and $\Phi\left(m_j-M_c\vert m_i\right)$ is on average larger than $\Phi\left(m_j-M_c\right)$. Indeed, if a previous earthquake has been detected, the probability to detect a subsequent larger one is larger compared to the case where no information about previously detected earthquakes is available.
At the same time, imposing $m_j \ge M_{\text{th}}$ and by considering increasing values of $M_{\text{th}}$, both detection functions $\Phi\left(m_j-M_c\vert m_i\right)$ and $\Phi\left(m_j-M_c\right)$ approach $1$, and $\Delta P(M,Y)$ approaches zero. Accordingly, the behavior of $E\left[\Delta P(M,Y)\right]$ in Fig.\ref{fig1}a cannot exclude the null hypothesis scenario where magnitudes are i.i.d variables in the presence of detection problems.

We now explore the prediction of the null hypothesis on the magnitude difference $q_i = m_{i+1} - m_i$, restricting to  $m_{i+1} > m_i$ and  $\delta y_i=\delta r_i <Y=d_0$.
The parameter $d_0$ is chosen sufficiently small to ensure that earthquake pairs are close enough in space, guaranteeing similarity in the distance to the stations necessary for their identification. Under this condition, since the earthquake $m_i$ has been detected, it implies that $m_i > M_c - 2\sigma$, and therefore $\Phi\left(m_j + M - M_c \vert m_i\right) \simeq 1$, for $M > 2\sigma$.
Accordingly, from Eq.(\ref{pcond2}), for $M >2\sigma$, $P(q_i> M\vert \Delta y_i<Y)=e^{-\beta M} K$, where $K$ is a constant given by $K=\int_{m_L}^{\infty}dm_i e^{-2\beta m_i} \Phi\left(m_i-M_c\right)$. After derivation, we therefore obtain, for $M>2\sigma$
\be
P(q_i=M \vert \Delta y_i<Y)=\beta K e^{-\beta M},  
\label{pdm4b}
\ee
showing that, even in the presence of detection problems, the distribution of magnitude differences $P(q_i)$ follows an exponential GR decay for $q_i>2 \sigma$, independently of the previous seismic history.
This behavior is observed in instrumental catalogs (Fig.\ref{fig1}b), where we find that $P(q_i)$ follows a pure exponential decay when $q_i$ is larger than a completeness threshold $\delta M_{c}=0.4$, for both SC and NC, consistent with Eq.(\ref{pdm4b}), with the same value of the exponent ($\beta=2.23\pm0.05$ in SC, $\beta=2.20 \pm0.05$ in NC)  extrapolated from the magnitude distribution $P(m_i)$ (Eq.(\ref{punfac})) when $m_i>M_c$ (Fig.\ref{fig1}b).

\begin{figure}[htp]
    \centering
    \includegraphics[width=9cm]{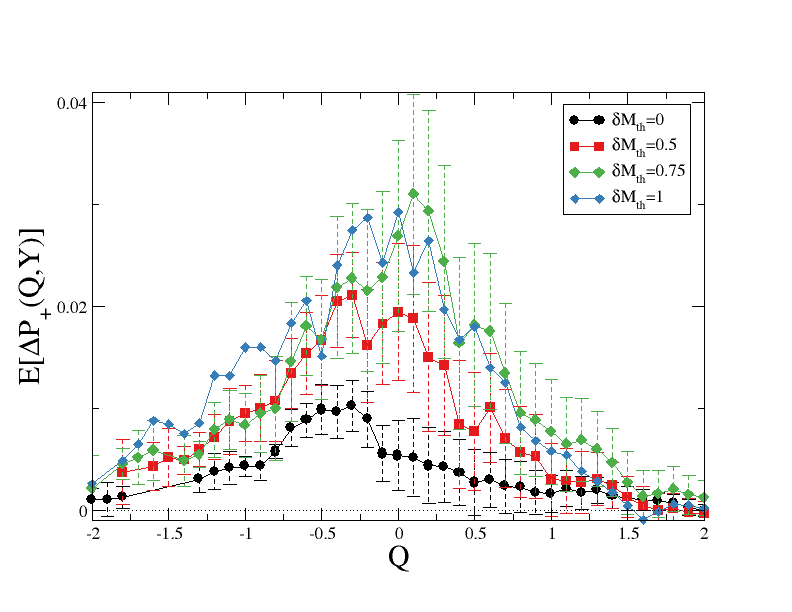}
    \caption{The quantity $E\left[\Delta P_+(Q,Y)\right]$ for the SC, where $\Delta P_+(Q,Y)$ is defined in Eq.(\ref{dp2}), with $\delta y_i=\delta t_i$ and $Y=1$ h. Different curves correspond to different thresholds $\delta M_{th}$. Error bars represent $2\Sigma(Q,Y)$ and are not shown for the curve with $\delta M _{th}=1$ to enhance readability.}
    \label{fig2}
\end{figure}

It is crucial to note that while detection issues introduce spurious correlations between $m_i$ and $m_{i+1}$, positive magnitude differences remain unaffected by such detection problems. Consequently, when $q_i > 2\sigma$, the null hypothesis predicts no correlations between $q_i$ and $q_{i+1}$. To delve deeper into this aspect, we select the subset of the catalog $\mathcal{D}$, denoted as $\mathcal{D}'$, which contains only those earthquakes satisfying both constraints $\delta r_i < d_0$ and $\delta m_i>0$, and therefore the $i$-th element of the catalog $\mathcal{D}'$ will be the vector ${q_i,t_i,x_i,y_i}$.
From the catalog $\mathcal{D}'$, we next obtain the catalog $\mathcal{DD}$, whose elements are the vectors $\delta q_i,t_i,\vec r_i$ with $\delta q_i=q_{i+1}-q_i$.
It is important to note that, in general, $\delta q_i \neq m_{i+2}-m_i$, and this equality only holds when $m_{i+2}>m_{i+1}>m_i$ and $dr_i$ and $dr_{i+1}$ are both smaller than $d_0$. We also generate several reshuffled catalogs $\mathcal{DD}_{\text{ran}}$, where the $i$-th element is the vector $\delta q^*_i,t_i,\vec r_i$ with $\delta q^*_i=q_{k}-q_i$, and $k$ is the index of a random earthquake in the catalog  $\mathcal{D}'$. We can define, $P(\delta q_i>Q \vert \delta y_i<Y)$, $\overline{P(Q,Y)}$, and $\Sigma (Q,Y)$ by simply replacing $m_i$ with $q_i$ and $M$ with $Q$ in all the definitions from Eq.(\ref{pcond}) to Eq.(\ref{pcond2}).
Accordingly, from Eq.(\ref{dp}), under the null hypothesis Eq.(\ref{punfac}), we obtain
\be
\begin{aligned}
& \Delta P_+(Q,Y)= P(\delta q_i>Q \vert \Delta y_i<Y)-P(\delta q^*_i> Q \vert \Delta y_i<Y) \\
& =\int_{0}^{\infty}dq_i \int_{q_i+Q}^{\infty}dq_j P(q_i)P(q_j) \Phi\left(q_i\right) \left [\Phi_q\left(q_j\vert q_i\right) -\Phi_q\left(q_j\right)\right],
\end{aligned}
\label{dp2}
\ee
where $\Phi_q\left(q_j\vert q_i\right)$ and $\Phi_q(q_j)$ represent the conditioned and unconditioned probabilities, respectively, of detecting a magnitude difference $q_j$. Correlations between $q_i$ and $q_{i+1}$ are induced by the term $\Phi_q\left(q_j\vert q_i\right)$. Nevertheless, in contrast to $\Phi\left(m_j-M_c\vert m_i\right)$ where the information that event $m_i$ has been detected strongly constrains the detectability of $m_{i+1}>m_i$, in this case, the information that the previous magnitude difference $q_i$ has been detected only weakly affects the probability $\Phi_q\left(q_j\vert q_i\right)$ to detect the next magnitude difference $q_{i+1}$. In fact, we expect that $\Phi_q\left(q_j\vert q_i\right) \simeq \Phi_q\left(q_j\right)$,
and $\Delta P_+(Q,Y)$ is expected to be very small under the null-hypothesis scenario. In particular, by considering only magnitude differences $q_i \ge \delta M_{th}$, as $\delta M_{th}$ increases, the detection probability $\Phi_q$ also increases and approaches $1$ for $\delta M_{th} \ge 2\sigma$. Consequently, we expect that $\Delta P_+(Q,Y)$ presents small values when $\delta M_{th}=0$, and decreases as $\delta M_{th}$ increases, vanishing when magnitude differences are detected with probability $1$ for $\delta M_{th} \ge 2\sigma$.

In the following we investigate the behavior of the mean value $E\left[\Delta P_+(Q,Y)\right]$ for the SC catalog. The main result is that $E\left[\Delta P_+(Q,Y)\right]$ exhibits a markedly different pattern (Fig.\ref{fig2}) compared to the one predicted by the null hypothesis (Eq.(\ref{dp2})). While data for $\delta M_{th}=0$ could be consitent with the null hypothesis Eq.(\ref{dp2}), as evidenced by $E\left[\Delta P_+(Q,Y)\right]$ being only slightly larger than $2\Sigma(Q,Y)$, the remarkable finding is that for larger values of $\delta M_{th}$, $E\left[\Delta P_+(Q,Y)\right]$ not only fails to decrease but rather increases. Particularly notable is that for $\delta M_{th} \ge 0.5$, significantly surpassing $\delta M_c \simeq 0.4$, when the variables $q_i$ remain unaffected by detection issues (Fig.\ref{fig1}b), the correlation between $q_i$ and $q_{i+1}$ becomes even more pronounced. The significant growth of $E\left[\Delta P_+(Q,Y)\right]$ as a function of $\delta M_{th}$  is notable until $\delta M_{th}=0.75$, after which no substantial increase is observed up to $\delta M_{th}=1$.
Results plotted in Fig.\ref{fig2} unequivocally demonstrate the presence of non-trivial correlations among magnitude differences. Consequently, the result derived from the null hypothesis of i.i.d. magnitudes, as expressed in Eq.(\ref{dp2}), can be confidently dismissed.

\begin{figure*}
    \centering
    \includegraphics[width=18cm]{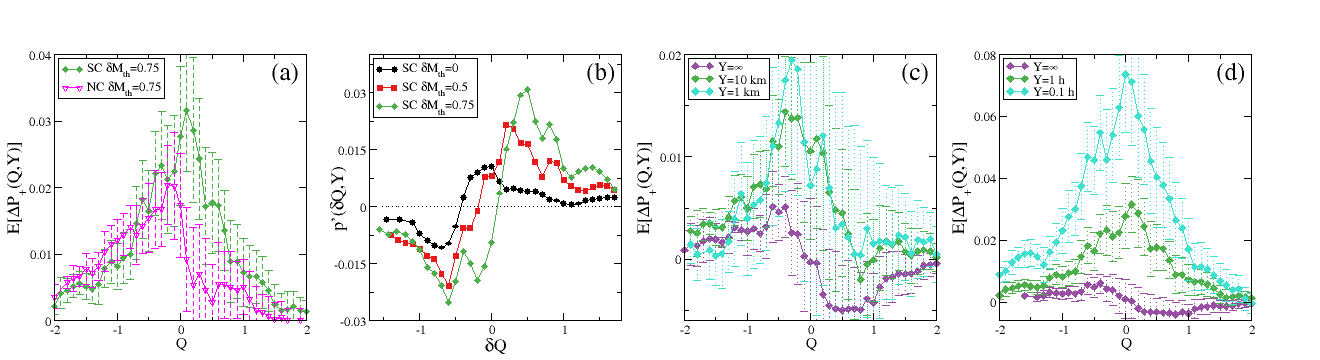}
    \caption{(a) The quantity $E\left[\Delta P_+(Q,Y)\right]$ for the SC and NC, with $\delta y_i=\delta t_i$ and $Y=1$h and $\delta M_{th}=0.75$. (b) The quantity $p'(\delta Q,Y)$ for  the SC, with $\delta y_i=\delta t_i$ and $Y=1$h  and different curves corresponding to different thresholds $\delta M_{th}$. (c) The quantity $E\left[\Delta P_+(Q,Y)\right]$ for the SC, with $\delta M_{th}=0.75$, $\delta y_i=\delta r_i$ and differnt values of $Y$. (c) The quantity $E\left[\Delta P_+(Q,Y)\right]$ for the SC, with $\delta M_{th}=0.75$, $\delta y_i=\delta t_i$ and different values of $Y$. In panels (a,c,d) error bars represent $2\Sigma(Q,Y)$.}
    \label{fig3}
\end{figure*}

We note that the results presented in Fig.\ref{fig2} were derived by considering only earthquakes with epicentral distances smaller than $d_0=20$ km as elements of the catalog ${\cal D}'$. Similar results were obtained for other choices of $d_0 \le 50$ km. Additionally, comparable outcomes were observed when analyzing the NC catalog (Fig.\ref{fig3}a). This suggests that correlations between magnitude differences appear to be a consistent characteristic of seismic catalogs, with patterns showing weak dependency on the geographic region.

In order to gain further insights into the mechanisms responsible for correlations between $q_{i+1}$ and $q_i$, we examine the influence of $Y$ on $E\left[\Delta P(Q,Y)\right]$ in Fig.\ref{fig3}. Specifically, in Fig.\ref{fig3}d, we consider $\delta y_i=\delta t_i$ and investigate decreasing values of $Y$, ranging from $Y=\infty$ to $Y=0.1$ hours. The results show that as $Y$ decreases, imposing that earthquake pairs $(q_i,q_{i+1})$ are closer in time, their correlation strengthens. 
In Fig.\ref{fig3}c, we conduct a similar analysis but with $\delta y_i=\delta r_i$, with $Y$ decreasing from $Y=\infty$ to $1$ km. Here, we observe that magnitude correlations become more pronounced as we transition from $Y=\infty$ to $Y=10$ km, but then stabilize, remaining roughly constant regardless of the specific value of $Y$. Consequently, magnitude correlations are stronger for earthquakes occurring closer in time, whereas they are only weakly affected by the spatial distance between earthquakes.

Further insights are provided by examining the derivative $p'(\delta Q,Y)=-\left.\frac{\partial E\left[\Delta P(Q,Y)\right]}{\partial Q}\right \vert _{\delta Q}$, which measures the difference in probability between the instrumental and reshuffled catalogs to observe $q_{i+1}=q_i+\delta Q$. The results (Fig.\ref{fig3}b) reveal that it is more probable for a magnitude difference to be followed by a larger one. Given that the inverse of the average value of $q_i$ coincides with $\beta$ \cite{VdE21}, the origin of magnitude correlations could be linked to a decreasing $\beta$ value over time during aftershock sequences. This observation aligns with the pattern identified by Gulia \& Wiemer \cite{GRTVEW18} from their analysis of $58$ main-aftershock sequences.
On the contrary, the scarcity of earthquakes preceding significant shocks poses a challenge in deriving conclusive insights from our analysis regarding the behavior of $\beta$ for foreshocks, and in comparing them with the predictive framework proposed in references \cite{GW19,GWV20}.

In summary, our findings unequivocally demonstrate the presence of non-trivial correlations between subsequent magnitude differences, suggesting potential modifications to be integrated into the ETAS model. One possible approach, as proposed by \cite{NOS19,NOS22}, is to consider two distinct $\beta$ values for aftershocks, $\beta \pm \delta \beta$, depending on the magnitude of the triggering mainshock. Alternatively, a time-varying $\beta$ value during the aftershock sequence could also be implemented \cite{GTWHS16}. This novel research direction holds promise for extracting valuable insights from the intricate patterns of seismicity, ultimately enhancing our ability to forecast seismic events and advancing our understanding of the mechanisms governing earthquake triggering.

\bibliographystyle{apsrev4-2}
\bibliography{../review}
\acknowledgments{ 
  E.L. and L.de A. acknowledge support from the MIUR PRIN 2022 PNRR  P202247YKL,
  C.G. acknowledges support from the MIUR PRIN 2022 PNRR P20222B5P9}

\end{document}